\input ppltexa.sty
\input epsf
\hsize 5.5 in \hoffset=0.5in
\vsize 8 in \def\pagesize{8.5 in} \voffset=0in
\def\Dscr{{\cal D}}
\def\Cscr{{\cal C}}
\def\Sscr{{\cal S}}
\def\ldotss{\ldots{}}
\let\eqv=\equiv
\def\bigglp{\biggl (}
\def\biggrp{\biggr )}
\let\mod\pmod 
\let\union\bigcup
\long\def\endpaper{\ifempty\refsfile\else\par\input\refsfile\fi\par\vfill\end}
\defrefsfile{correlref.sty}
\def\refstyle{B}\def\seccase{L}\def\secnumbering{A} 
\defsections  {\intro \map \trapping \results
	       \hard \discuss \conclusions }
\defappendices{\mud \numer \markov }
\endpreamble
\hbox to \the\hsize{\hss \hss
			\hbox to 0 pt{\hss PPPL--1938 (Nov.\ 1982)}}
\hbox to \the\hsize{\hss \hss
	\hbox to 0 pt{\hss Physica \vol{8D}(3), 360--380 (Sept.~1983)}}
\vskip 0.1 in
\hbox{\titl LONG-TIME CORRELATIONS IN}
\vskip 5 pt
\hbox{\titl THE STOCHASTIC REGIME}
\vskip 10 pt
\hbox{\hskip20pt Charles F. F. KARNEY}
\vskip 2 pt
\hbox{\hskip20pt Plasma Physics Laboratory, Princeton University,}
\hbox{\hskip20pt Princeton, New Jersey 08544, USA}
\vskip -10 pt
\abstract

The phase space for Hamiltonians of two degrees of freedom is usually
divided into stochastic and integrable components.  Even when well
into the stochastic regime, integrable orbits may surround small
stable regions or islands.  The effect of these islands on the correlation
function for the stochastic trajectories is examined.  Depending on the
value of the parameter describing the rotation number for the
elliptic fixed point at the center of the island,
the long-time correlation function may decay as $t^{-5}$ or
exponentially, but more commonly it decays much more slowly (roughly as
$t^{-1}$).  As a consequence these small islands may have a profound
effect on the properties of the stochastic orbits.  In particular,
there is evidence that the evolution of a distribution of particles is
no longer governed by a diffusion equation.

\vfill\noindent
This paper was reprinted in {\it Hamiltonian Dynamical Systems},
edited by R. S. MacKay and J. D. Meiss (Adam-Hilger, Bristol, 1987),
pages 585--605.
\par\eject
\section \intro.  Introduction

Many important problems in physics are described by Hamiltonians of two
degrees of freedom.  Examples are the motion of a charged particle in
electro\-static waves, the motion of a charged particle in various
magnetic confinement devices, the acceleration of a particle bouncing
between a fixed and an oscillating wall, the wandering of magnetic
field lines, etc.  In such systems, there may be a parameter $k$
which governs the strength of the coupling between the two degrees of
freedom.  When $k$ is zero (no coupling), the system is
integrable.  As $k$ is increased, some of the integrable
trajectories disappear, and the motion becomes stochastic.
Eventually, when the $k$ is large, nearly all of phase space is
occupied by stochastic trajectories.  These properties are nicely
illustrated by the standard mapping\ref\chirikov.  Although this mapping
is an idealization, it accurately portrays many of the properties of
real systems.

When $k$ is small, various perturbation theories are available to
describe the trajectories approximately.
On the other hand, in the highly stochastic
regime ($k$ large) certain statistical quantities may be analytically
determined by assuming that the motion is ergodic and
that the correlation time is short.  In the case of the standard mapping,
this allows simple determinations of quantities such as the diffusion
coefficient\ref{\chirikov, \rw}\ and the KS-entropy\ref\chirikov.

The purpose of this paper is to examine more critically the assumption
that the correlation time is short in the stochastic regime.  Our
interest in this problem was triggered by studies of the diffusion
coefficient for the standard mapping,
$$r_t-r_{t-1}=-k\sin\theta_{t-1},
\qquad \theta_t-\theta_{t-1}=r_t.\eqn(\en\sm)$$
Far above the stochasticity threshold $k\grgr 1$, the diffusion
coefficient defined as
$$\Dscr=\lim_{t\to\infty}{\ave{(r_t-r_0)^2}\over2t},$$
where the average is over some appropriate ensemble, is given by
assuming that $\theta$ is a random variable in the equation for $r$.
This is equivalent to assuming that the correlation function is
proportional to a delta function and
gives the ``quasi-linear'' result $\Dscr/\Dscr_{\subrm{ql}}=\quarter
k^2$.  Including the correlations out to short times (about $t=4$) gives
corrections to the diffusion coefficient reported by Rechester and
White\ref\rw\ which can enhance the diffusion coefficient by as much as
a factor of about two.  However, a numerical determination\ref\krw\ of
the diffusion at $k=6.6$ gave $\Dscr/\Dscr_{\subrm{ql}} \sim 80$.  At this
value of $k$ there is an island (``accelerator mode'') present in the
stochastic sea.  Although the orbits used to compute $\Dscr$ were all in
the stochastic region, they were able to wander close to the island and
stay close to it for a long time.  This introduces long-time
correlations into the motion and accounts for the anomalously large
value of $\Dscr$ observed.

The appearance of islands in the stochastic regime is not at all
unusual.  This may be seen from
Sinai's estimate\ref\chirikov\ for the number $\nu(T)$ of periodic
orbits of period $\leq T$:  $\nu(T) \sim \exp(hT)$ for $T\to\infty$ where
$h$ is the KS-entropy.  Now the majority of these periodic orbits are
unstable.  However as $k$ is increased, $h$ generally increases ($h
\approx\log(\half k)$\ref\chirikov\
for the standard mapping) and hence new periodic
orbits must appear.  Generally a tangent bifurcation is responsible for
the appearance of the new periodic orbits.  At the tangent bifurcation, a
pair of stable and unstable periodic orbits is born.  The stable orbit
gives rise to other longer periodic orbits as $k$ is increased and
eventually becomes unstable.  Thus
as we increase $k$, many small islands appear via tangent bifurcations,
survive for some interval of $k$, and finally disappear (usually
through period doubling).  If we pick a particular value of $k$, it is
not clear that there will necessarily be any islands present.  However,
we may speculate that at some arbitrarily close value of $k$, there
will be some islands.

It may be objected that the large effect seen in the standard mapping
arises because the islands are accelerator modes\ref\chirikov\ and
that such islands
are a rather special feature of the standard mapping.  While accelerator
modes are the only islands which will contribute significantly to the
force (i.e., acceleration) correlation function and hence to the
diffusion coefficient, any islands will contribute to the correlation
of some functions on phase space.  The results of this study will be
applicable to systems like the Fermi map which have no accelerator
modes.  This study also contributes to the understanding of the
more general problem of motion in a divided phase space.

In this paper, we wish to examine more closely the effect these islands
have on a stochastic trajectory.  As far as determining the effect on
the correlation function, this involves determining how ``sticky'' the
island is.  Given that the stochastic trajectory comes within a certain
distance of the boundary of the island, how long do we expect it to stay close to the
island?  This approach is inspired by work of Channon and
Lebowitz\ref\channon\ on the correlations of a trajectory in the
stochastic band trapped between two KAM surfaces in the H\'enon map.  
Similar work has been carried out on the whisker map by Chirikov
and Shepelyansky\ref\chirikova.  This work is being extended by B. V.
Chirikov and F. Vivaldi.

Since we concentrate only on the behavior close to the island, this
approach may be characterized as a local one.  This should be compared
with Fourier transform methods\ref\rw, which are global and are not well
suited to the description of localized phenomena.  For instance, Meiss
{\it et al.}\ref\meiss\ attempted to use such methods to compute the
long-time correlations for the standard mapping, and they found poor
agreement with numerical experiments whenever islands were present.

The paper is organized as follows.  In section \map, we derive a
canonical mapping which describes the behavior near an island.  Next
(section \trapping), we define the trapping statistics which describe
how sticky the island is.  The results for the trapping statistics are
given in sections \results\ and \hard.  In section \discuss,  we show
how to apply these results to obtaining the correlation function.  The
results are discussed in section \conclusions.

\section \map.  Derivation of mapping.

Far into the stochastic regime for a general mapping,
the islands which appear via tangent
bifurcations are very small and exist only for a small interval in
parameter space.  This allows us to approximate them by
a Taylor expansion in both
phase and parameter space about the tangent bifurcation point
retaining only the leading terms.  This was carried out in ref.\ \krw\
where the resulting mapping was reduced to a canonical form
$$Q:\qquad y_t-y_{t-1} = 2(x_{t-1}^2-K)\eqv g(x_{t-1};K),
\qquad x_t-x_{t-1}=y_t.\eqn(\en\qu)$$
Here $K$ is proportional to $k-k_{\subrm{tang}}$ ($k_{\subrm{tang}}$ is
the parameter value where the tangent bifurcation take place) and $x$
and $y$ are related to the original phase space coordinates by a smooth
transformation.  The mapping $Q$ represents an approximation of the
general mapping close to the point of tangent bifurcation.
For $K<1$, $Q$ has no periodic orbits.  At $K=0$,
there is a tangent bifurcation when an unstable fixed
point appears at $x=y=0$.  (This is not a hyperbolic point since its
stability is determined by the quadratic terms in the mapping.)
For $0<K<1$, this fixed point splits into a pair of stable (elliptic)
and unstable (hyperbolic)
fixed points located at $(x,y)=(\mp \sqrt K,0)$, respectively.  The
elliptic fixed point is usually surrounded by integrable trajectories
(KAM curves) which define a stable region (the
island) in which the motion is bounded.  An
example of island structure is shown in fig.\ \exam\ for
$K=0.1$ (the value of $K$ at which extensive numerical calculations have
carried out).  At
$K=1$ the stable fixed point becomes unstable and gives rise to a
period-2 orbit via a period-doubling bifurcation.  At
$K=1.2840$ the period-doubling sequence accumulates\ref\gmvf, and
at this point (or shortly hereafter\ref\mackay) the area of the
stable regions becomes very small.  For $0<K<1$,
the mapping $Q$ may be transformed to the H\'enon quadratic map\ref\henon,
the parameter $K$ being related to H\'enon's $\cos\alpha$ by
$$\cos\alpha=1-2\sqrt K$$
($\alpha$ is the mean angle of rotation for points close to the stable
fixed point).

Referring again to the islands shown in fig.\ \exam, consider a
particle which at $t=0$ is close to, but outside, the islands.  (We
often speak of an orbit in terms of the position of a particle whose
equation of motion is given by $Q$.)  Initially,
the particle will stay close to the islands; however as we let
$t\to\pm\infty$, we find $(x,y)\to(\infty,\pm\infty)$.  It is just such
trajectories we are interested in, because they correspond to particles
in the stochastic region of the general mapping approaching the islands,
staying there for some time (and contributing to long-time
correlations), and then escaping back to the main part of the stochastic
region.

What we want to do is to follow such trajectories numerically and see
how long they stay close to the islands.  However, we need some method of
fairly sampling these trajectories.  ``Fairly'' means that we should
sample them in the same way that a stochastic trajectory of a general
mapping does.  Since the stochastic trajectory is ergodic over the
connected stochastic region of phase space, we must sample trajectories
in the same way; i.e., we must ensure that the superposition of all the
sampled trajectories covers the region outside the islands uniformly.

We achieve this by changing $Q$ so that the phase space is compact.
This may be accomplished by replacing $g(x;K)$ in $Q$ by the periodic
function
\def\xmax{x_{\subrm{max}}}\def\xmin{x_{\subrm{min}}}\def\qstar{Q^\ast}
$$g^\ast(x;K)\eqv\case{g(x;K)&\for \xmin\leq x <\xmax,\cr
		g(x\pm L;K)&\hbox{otherwise},\cr}\eqn(\en\gstar)$$
where $L=\xmax-\xmin>0$.  The resulting map will be called $\qstar$.  If
$\xmin$ and $\xmax$ are chosen to span the region where there are
islands in $Q$, then $\qstar$ obviously will contain the same islands.
Furthermore the motion close to the islands will be the same.  By
replacing $g$ by $g^\ast$, the whole of phase space becomes periodic in
the $x$ and $y$ directions with period $L$.  The motion can be treated
as though it were on a torus.  A particle which starts near the
island will, as with $Q$, spend some time close to the island.
But when it moves away from the island it no longer goes to infinity, but
rather it loops around the torus and has another chance to approach the
islands.  Since $\qstar$ is area-preserving, this single orbit will
ergodically cover the region outside the islands in the desired manner.
Examples of such orbits are shown in fig.\ \exa\ for the same
parameters as for fig.\ \exam.  (One embarassing feature of $\qstar$ is
that new islands are introduced.  They are in fact accelerator modes---a
particle inside one of them loops around the torus in either the $x$ or
$y$ directions.  As discussed in appendix \mud, these islands do not
effect the statistics for the long trapping times.)

One useful way of looking at $\qstar$ is as a magnification of a small
region near a tangent bifurcation in the general mapping.  The
difference is that once the trajectory leaves the vicinity of the
islands, it is immediately re-injected on the other side of the
islands.  In the general map, the trajectory will spend some long time,
which depends on the ratio of the size of the islands to the total
accessible portion of phase space, in the stochastic sea before coming
back to the vicinity of the islands.

Assuming that the long-time behavior of stochastic orbits is dominated
by the region close to the islands, there are two advantages to
reducing the problem to a study of $\qstar$.  Firstly, since $\qstar$
describes the behavior of most islands far into the stochastic regime,
the properties of many mappings may be treated by looking at a special
mapping $\qstar$ which depends only on a single parameter $K$.  The
second advantage is that the properties of orbits close to the islands
may be studied much more efficiently because there is no need to follow
orbits while they spend a long and uninteresting time far from the
islands.

\section \trapping.  The trapping statistics.

The prescription for numerically determining how sticky is the island
system in $Q$ for a given $K$ is to pick trajectories outside the
islands in $\qstar$ and to iterate the mapping many times.  The mapping
is performed on the torus, i.e., $x$ and $y$ are reduced to their base
intervals $[\xmin,\xmax)$ and $[-\half L,\half L)$ after each iteration.
However, we keep track of when an orbit moves off one edge and
re-appears at the opposite edge.  The orbit is then divided at those
points when the orbit looped around the torus, and the lengths of the
resulting orbit segments are recorded.  The main result of such a
calculation are then the {\it trapping statistics} $f_t$ which are
proportional to the number of orbit segments which have a
length of $t$.

Suppose that the total length of the orbit is $T$ and $N_t$ is the
number of segments of length $t$.  If $T$ is so large that we can ignore
partial segments at the ends of the orbit (this problem is examined
below), then we have $\sum tN_t = T$; the total number of segments is
$N=\sum N_t$.  The trapping statistics are defined by $f_t=N_t/T$ and
are therefore normalized so that $\sum tf_t=1$.  The mean length of the
orbits is given by $\alpha= 1/\sum f_t$ ($= T/N$).  The probability
that a particular segment has length $t$ is $p_t=\alpha f_t$
($=N_t/N$).  If an arbitrary point is chosen in the orbit, then $tf_t$
is the probability that this point belongs to a segment of length $t$
and $f_t$ is the probability that it belongs to the beginning, say, of
a segment of length $t$.

Three factors effect the measurement of $f_t$.  They
are (a) the presence of the spurious accelerator modes, (b) the choice
of $\xmin$ and $\xmax$, and (c) the total length $T$ of the trajectory used
to measure $f_t$.  The first two items only effect $f_t$ for small $t$
(apart from an overall normalization).  In order to account for
the last item, we define $f_t$ by $N_t/(T+1-t)$ (rather than $N_t/T$).
This accounts for the fact
that we are less likely to observe orbit segments whose length is
close to $T$.  All these
points are discussed in detail in appendix \mud.

The survival probability
$$P_t=\sum_{\tau=t+1}^\infty p_\tau\eqn(\en\pdef)$$
is the probability that an orbit beginning in a segment at $t=0$
is still trapped in
the same segment at time $t$.  Note that $P_0=1$ as required.  This is
the quantity studied in refs.\ \channon\ and \chirikova.
The correlation function
$$C_\tau=\sum_{t=\tau}^\infty (t-\tau)f_t
=\sum_{t=\tau}^\infty P_t/\alpha\eqn(\en\cdef)$$
is the probability that a particle is trapped in the same segment at
two times $\tau$ apart.  Again, we have $C_0=1$.

There are two other ways of interpreting $C_\tau$.  If we start many
particles at positions uniformly distributed in the stochastic sea of
$\qstar$ (i.e., in the dark region of fig.\ \exa), then $C_\tau$ gives
the fraction of particles remaining in the $L\times L$ square after
$\tau$ iterations of $Q$ (rather than $\qstar$).  Alternatively,
consider a drunkard who executes a one-dimensional random walk with
velocity $v=dr/dt=\pm1$.  The direction of each step is chosen randomly,
while the durations of the steps are chosen to be the lengths of
consecutive trapped segments of $\qstar$.  Then for integer $\tau$,
$C_\tau$ is just the usual correlation function for such a process,
i.e., $\ave{v_tv_{t+\tau}}_t$.  The behavior of this random-walk process
is similar to the behavior of an orbit in the general mapping when
two accelerator modes with opposite values of the acceleration are
present.  (This is the case with the first-order accelerator modes for
the standard mapping.)  In section \discuss, we will show how $C_\tau$
is related to the correlation function for the general mapping.

A diffusion coefficient may be defined by
$$D=\half C_0+\sum_{\tau=1}^\infty C_\tau=\sum \half t^2 f_t.\eqn(\en\ddef)$$
This gives the diffusion rate for the drunkard in the random-walk
problem above.  It is also related to the diffusion coefficient for
the general mapping (see section \discuss).

Since $\qstar$ must be iterated many times to provide good statistics
for $f_t$ for large $t$, extraordinary steps were taken to ensure that
the numerical program was fast and reliable.  The time for one iteration
on a Cray--1 is a little less than 75 ns.  One way that the code was
made more reliable was by doing the arithmetic in fixed-point (as
opposed to floating-point) notation.  The numerical mapping is then
one-to-one which is the discrete counterpart of area-preserving.
(Floating-point realizations of mappings are typically many-to-one.)
This precludes the possibility of an orbit, which starts far from the
island, approaching the island and becoming permanently trapped near
the island.  Even though such behavior is forbidden for an
area-preserving mapping, such behavior may be observing with a
floating-point realization.  Details of the numerical methods are given
in appendix \numer.

\section \results.  The results for small \uppercase{$K$}.

We begin by considering the cases where $K$ is small or zero.  In this
case the mapping equations are nearly integrable and this enables us to
derive approximate analytic expressions for $f_t$.
Figure \fsmall\ shows $f_t$ for
$K=-10^{-4}$, $0$, and $10^{-4}$.  Three types of behavior are seen for
$t\to\infty$:  a cutoff distribution $f_t=0$ for $t>t_{\subrm{max}}
\approx 50$, a rapid algebraic decay $f_t\sim t^{-7}$, and an exponential
decay $f_t\sim \exp(-0.1t)$.

The easiest case to begin with is $K=0$.  The method for deriving $f_t$
analytically consists of computing the length of the trajectory through
some point and then assigning some probability that this trajectory
will be chosen.  The first part of the calculation has been carried out
by Zisook\ref\zisook.  We repeat the calculation here to establish the
method for other cases.

When $K=0$, we are exactly at the tangent bifurcation point.  There are
no islands in this case, but trajectories can still spend arbitrarily
long near the fixed point at $(x,y)= (0,0)$.  Near this point
$x_t-x_{t-1}$ and $y_t-y_{t-1}$ are small.  We therefore rescale $x$,
$y$, and $t$ with $x=\epsilon^\alpha X$, $y=\epsilon^\beta Y$,
$t=\epsilon^{-1} T$ where $\epsilon$ is small.  The mapping $Q$
becomes
$$\eqalign{{Y(T)-Y(T-\epsilon)\over\epsilon}
&=\epsilon^{2\beta-\alpha-1} 2X^2(T-\epsilon),\cr
{X(T)-X(T-\epsilon)\over\epsilon}
&=\epsilon^{\alpha-\beta-1} Y(T).\cr}$$
Choosing $\alpha=3$ and $\beta=2$ and replacing the left hand sides by
derivatives, we obtain
$${dY\over dT}=2X^2,\qquad 
{dX\over dT}=Y.$$
These are Hamilton's equations (with $X$ and $Y$ being
conjugate position and momentum coordinates) for the Hamiltonian
$$H=\half Y^2-\fract2/3 X^3.\eqn(\en\ham)$$
Curves of constant $H$ in the $(X, Y)$ plane give the
trajectories, examples of which are shown in fig.\
\level(a).  We define the trapping time as the time it takes to traverse
one of these
curves from $Y=-\infty$ to $\infty$.  (This Hamiltonian gives
escape to infinity in a finite time.  The time taken for a particle to
escape to infinity in $Q$ is infinite, but very weakly so.  The
particle reaches $y$ from $y=0$ in roughly $\log\log y$ steps for $y$
large.) Since ${dX/dT}=Y=\sqrt{2(H+\fract2/3
X^3)}$, the trapping time may be written as
$$T(X_0)=2\int_{X_0}^\infty {dX\over
\sqrt{\fract4/3(X^3-X_0^3)} },$$
where $X_0=-(\fract3/2H)^{1/3}$ is the $X$ intercept
of the trajectory with $Y=0$.
Performing the integration gives
\def\nthroot#1#2{\rlap{\raise5pt\hbox
{\hskip2.5pt$\scriptscriptstype#1$}}\sqrt{#2}}
$$T(X_0)={1\comb\{\}\sqrt3}\times4.207 \abs{X_0}^{-1/2},\quad
\hbox{for }X\grls0.$$
(The numbers here may be written in terms of incomplete elliptic
integrals.)

This completes the computation of the trapping time.  We now assign
weights to each trapping time by requiring that particles
spend equal times in equal areas of phase space.  Let $A(X_0)$ be
the area between the trajectory passing through the origin and that
passing through $(X_0,0)$ in fig.\ \level(a).  Since $Y\sim X^{3/2}$, scaling
invariance gives $A(X_0) = A(1) X_0^{5/2}$.  (This procedure needs to be
carried out separately for positive and negative $X_0$.  However, the
scaling relations are the same in the two cases.)  Parameterizing in
terms of the trapping time $T$ gives $A(T)\sim T^{-5}$.  The fraction of
particles which are trapped for times between $T$ and $T+dT$ is
proportional to the differential area $dA(T) \sim T^{-6}dT$.  Finally, we
divide by $T$ to give $T^{-7}dT$ as the number of orbit segments of
lengths in this range.  In unscaled variables, we have $f_t\sim t^{-7}$
which is valid for $t$ large.  The correlation function has the
asymptotic form $C_\tau\sim \tau^{-5}$.

It is interesting to enquire what the asymptotic behavior for $f_t$
would be if $g$ in (\map)
were a higher-order polynomial in $x$.  If $g(x;0)=x^m$,
with $m>1$, we can repeat the above calculation and find that
$$f_t\sim {1\over t^{(3m+1)/(m-1)}}.$$
Since $f_t$ has a finite second moment, $D$ always exists.
(For $m=1$, we find an exponential decay of $f_t$.  This corresponds to
the case $K>0$ discussed below.)

When $K$ is small and negative, there are no periodic orbits.  The particle
can spend only a bounded time close to the origin.  Defining scaled
variables as before, together with $K=-\epsilon^4$, gives differential
equations which are derivable from the Hamiltonian
$$H=\half Y^2-\fract2/3 X^3-2X.\eqn(\en\hama)$$
The trapping time is now
$$T(X_0)=\int_{X_0}^\infty {dX\over
\sqrt{\fract1/3X^3+X-\fract1/3X_0^3-X_0} },\eqn(\en\tneg)$$
\def\tmax{T_{\subrm{max}}}%
where $X_0$ is the $X$ intercept of the trajectory with
$Y=0$, i.e., it is the real root of $H+\fract2/3X_0^3+2X_0=0$.
This is plotted as a function of $X_0$ in fig.\ \txzero(a).  We see it
attains a maximum value of $\tmax=5.1454$ at $X_0=-0.5536$.  In unscaled
variables this means that the maximum trapping time is $t_{\subrm{max}}
=5.1454\abs
K^{-1/4}$ and that this trapping time is attained by particles passing
through $(x,y)=(-0.5536\sqrt{\abs K},0)$.  The $\abs K^{-1/4}$ scaling
of the maximum trapping time has been derived by Zisook\ref\zisook.

To assign probabilities to the various trapping times, we define
$A(X_0)$ as the area between the trajectory passing through $(0,0)$ and
that passing through $(X_0,0)$.  Since
$Y=2\sqrt{\fract1/3X^3+X-\fract1/3X_0^3-X_0}$, we obtain
$$A(X_0)=4\int_0^\infty\sqrt{\fract1/3X^3+X}\,dX
-4\int_{X_0}^\infty\sqrt{\fract1/3X^3+X-\fract1/3X_0^3-X_0}\,dX.$$
(The two integrals need to be done to together to get a finite
answer.) Differ\-entia\-ting gives
$$dA(X_0)/dX_0=2(X_0^2+1)T(X_0).$$
The number of orbit segments of length $T$ is then proportional to
$$F(T)\eqv
{1\over T}\left.{dA(X_0)\over dX_0}\right/\left|{dT(X_0)\over dX_0}\right|
=\sum{2(X_0^2+1)\over\abs{T^\prime(X_0)}}.\eqn(\en\fneg)$$
The right hand side is written as a function of $X_0$.  This is
converted to a function of $T$ by inverting $T(X_0)$.  Since this gives
a double-valued function, the two branches must be summed over as
indicated by the summation sign.  In unscaled variables we have
$$f_t\sim F(\abs K^{1/4}t)$$
for $t=O(\abs K^{-1/4})$.  The function $F(T)$ is plotted in fig.\
\txzero(b).  For $T\lsls\tmax$, we have $F(T)\approx 6.205\times10^5T^{-7}$
while for $T\approx \tmax$, $F(T)\approx 3.024 (\tmax-T)^{-1/2}$.
The correlation function is given by the second integration of $f_t$ so
that for $\tau\approx t_{\subrm{max}}$ we have
$C_\tau\sim (t_{\subrm{max}}-\tau)^{3/2}$ for $\tau\leq t_{\subrm{max}}$.

Figure \txzero(b)
should be compared with fig.\ \fsmall(a).  The effect of the
singularity in $F(T)$ at $\tmax$ is evident, although its position
isn't quite right.  Furthermore, the decay of $f_t$ for smaller times is
somewhat slower (approximately as $t^{-6}$) than for $F(T)$.  These
discrepancies arise because we are not far enough into the asymptotic
regime since $\epsilon$ is not very small (0.1).  The formula
$t_{\subrm{max}}=5.1454\abs K^{-1/4}$ may easily be verified for
smaller values of $\abs K$.  The $T^{-7}$ behavior of $F(T)$ has of
course the same origin as that of $f_t$ for $K=0$.  However the
numerical results for $K=0$ given in fig.\ \fsmall(b) show that it is
not attained until about $t\sim 100$.  When $K=-10^{-4}$, $t_{\subrm{max}}$
is only about $50$, and there is no interval in which the
$t^{-7}$ behavior is exhibited.

Finally we turn to the case $K>0$.  As with $K<0$, we can approximately
derive the motion from the Hamiltonian
$$H=\half Y^2-\fract2/3 X^3+2X,\eqn(\en\hamb)$$
where we have used the same scaled variables as previously except that
$K=\epsilon^4$.  The trajectories for this Hamiltonian are given in
fig.\ \level(b).  There is single island centered at the fixed point at
$(X,Y)=(-1,0)$ and the island extends all the way to the separatrix
emanating from the unstable fixed point at $(X,Y)=(1,0)$.  This is an
idealization because for the mapping there is a stochastic band close
to the separatrix.  However, for small $K$ this band is very thin, and
the island does have quite a ``clean'' outer edge.

We may carry out the analysis given for $K<0$ with appropriate changes
to obtain $f_t$ analytically.  However, we may avoid doing a lot of
tedious integrals by concentrating only on the long-time behavior of
$f_t$.  For shorter times, namely for $100\lsapprox t\lsls K^{-1/4}$, we
expect that $f_t\sim t^{-7}$ because the relevant trajectories never get
close to the island and the Hamiltonian (\hamb) approximately reduces
to that for $K=0$ (\ham).
Since only the region close to the unstable fixed point contributes to
$f_t$ for large $t$, we need only consider this region.  In addition,
we should distinguish those trajectories which encircle the island and
so encounter the region near the fixed point twice from those which do
not and only encounter this region once.  For a given distance from the
fixed point, the trapping time of orbits in the former category will be
about twice as long as those in the latter category.

Near any hyperbolic point there exist coordinates $(\xi,\eta)$ in which
the mapping may be written as
$$\xi_t=\lambda \xi_{t-1},\qquad \eta_t=\lambda^{-1} \eta_{t-1},$$
where $\lambda$ is the Lyapunov number at the fixed point.  The
trajectories lie on hyperbolae of the form $\xi\eta=\pm \xi_0^2$.  Since the
island encircling trajectories dominate the long-time behavior of
$f_t$, we define the trapping time $t(\xi_0)$ as the twice the time it
take to get from $\eta=1$ to $\xi=1$.  (The factor of two accounts for the
two encounters with the fixed point.)  We find
$$t(\xi_0)=-4\log \xi_0/\log\lambda,\qquad 
dt(\xi_0)/d\xi_0=-4/(\xi_0\log\lambda).$$
The area $A(\xi_0)$ between the hyperbola, the axes, and the lines $\xi=1$
and $\eta=1$ is
$$A(\xi_0)=\xi_0^2(1-2\log \xi_0).$$
The trapping statistics are then given by
$$f_t\sim {1\over t}{dA\over dt}\sim \exp(-\half\log\lambda\, t).$$

For the fixed point at $(0,\sqrt K)$,
$$\lambda=1+2\sqrt K+2\sqrt{\sqrt K +K}\approx 1+2K^{1/4},
\qquad f_t\sim \exp(-K^{1/4}t).$$
$C_\tau$ behaves in the same way.
For $K=10^{-4}$, the decay rate should be about $0.1$, which is
indeed what was observed in fig.\ \fsmall(c).  Similar agreement is
seen at $K=10^{-3}$ and $10^{-2}$.  However at $K=0.1$, the central
island has shed a chain of sixth-order islands and the foregoing
analysis does not apply.  This case is discussed in the next section.

\section \hard.  The results for \uppercase{$K=0.1$}.

We have measured $f_t$ for $K$ between 0 and 1.3 at intervals of 0.05,
and at most of the values of $K$ a slow algebraic decay of $f_t$ is
seen.  A representative case is $K=0.1$, whose trapping statistics are
given in fig.\ \expense(a), which illustrates the slow decay for very
long times $t\sim 10^7$.  Also given in fig.\ \expense\ are $P_t$, $C_\tau$,
and $\alpha\eqv-d\log C_\tau/d\log\tau$ (thus locally
$C_\tau\sim \tau^{-\alpha}$).  This last plot shows the power at which
$C_\tau$ decays varying between about $\fract1/4$ and $\fract3/2$.

A glance at fig.\ \exa\ shows the origin of this behavior.  The central
island is surrounded by a chain of sixth-order islands.  Around each of
these islands are several other sets of islands.  This picture repeats
itself at deeper and deeper levels.  A particle which manages to
penetrate into this maze can get stuck in it for a long time.

For $\tau\lsapprox10^4$, fig.\ \expense(d) gives
$\alpha\approx\fract1/4$.  Correspondingly we have $P_t\sim t^{-p}$ where
$p=1+\alpha\approx 5/4$.  This is close to the asymptotic ($t\to\infty$)
result found in ref.\ \chirikova\ for the whisker map, in which
$\ave p\approx 1.45$.  This is another indication that the behavior of a
Hamiltonian with a divided phase space has ``universal'' properties.
However, in our case, $\alpha$ shows some strong variations beyond $\tau
\approx10^4$ where $C_\tau$ ``steps down'' (e.g., between $10^4$ and
$3\times10^5$).  This means that the asymptotic form of $C_\tau$ is
very difficult to determine numerically.

The diffusion coefficient $D$ is given by the summation of $C$ and is
approximately $6400\pm800$.  The error is estimated by calculating $D$
separately for subsets of the orbits sampled.  Unfortunately, because
a few very long orbit segments have such a large effect on $D$,
the individual observations of $D$ come from a highly skewed
distribution and the error may be severely underestimated.
We will try to get a idea of the error by asking
what behavior is possible for $f_t$ for $t_a=10^8<t<
2\times10^9=t_b$ ($t_b$ is the length of the orbits used to compute
$f_t$ in fig.\ \expense).  Since no segments were observed in this range,
we have
$$\int_{t_a}^{t_b}M(t_b-t)f_t \,dt\lsapprox 1,$$
where $M=1600$ is the number of orbits used.  This just says that the
expected number of orbit segments in this range of $t$ is less than 1 (see
appendix \mud).  Suppose that $f_t=A (t/t_a)^{-(2+\alpha)}$ where
$\alpha$ is the exponent at which $C_\tau$ decays and
$A$ is the value of $f_t$ at $t=t_a=10^8$.  For
$t_b\grgr t_a$ and $\alpha>0$, we can evaluate the integral to give
$MAt_at_b/(1+\alpha)$ approximately.  If we take
$A=10^{-20}$ (this value was estimated from fig.\ \expense(a)), we find
$\alpha\grapprox2$.  In the case of the slowest decay,
$\alpha=2$, the portion of 
$f_t$ between $t_a$ and $\infty$ would increase $D$ by about a factor of
$2$ over the value given above.  If $f_t$ takes another
step down near $t=10^8$, then $A$ might be smaller and smaller values
of $\alpha$ would be possible and the maximum error in $D$ would be
larger.  For instance with $A=0.3\times10^{-20}$,
then all values of $\alpha<0$ are consistent with the
numerical observations.  Since $C_\tau$ sums to infinity for
all $\alpha\le1$, $D$ may well be infinite!

If $D$ is indeed infinite, we would wish to know how a group
of particles spreads with time.  We again consider
the drunkard's walk based on $\qstar$ which was introduced in section
\trapping.  The second moment of $r$ is related to the correlation
function by
$$S_t\eqv\ave{(r_t-r_0)^2}=tC_0+2\sum_{\tau=1}^t(t-\tau)C_\tau.
\eqn(\en\spreaddef)$$
This is plotted in fig.\ \spread(a), using the data of fig.\ \expense.
For $t\lsapprox10^4$, $S_t$ grows somewhat faster than $t^{3/2}$ (see
fig.\ \spread(b)) and even until $t\approx10^7$, $S_t$ is growing
significantly faster than linearly.  Beyond $10^7$, the numerical data
shows a convergence to a linear rate; but this is merely because no
segments longer than about $6\times10^7$ were observed.  For
$t\to\infty$, $S_t$ grows as $t^{2-\alpha}$, assuming that the exponent
$\alpha$ at which $C_\tau$ decays asymptotically is less than
$1$.  If the diffusion coefficient is estimated from $D_t=\half S_t/t$,
then $D_t$ grows with $t$ as shown in fig.\ \spread(c).

When applying these results to the general mapping, we will need
to know the area $b_2$ occupied by the stochastic trajectories (the dark area
in fig.\ \exa).  This was calculated by dividing the phase space into
$1024\times1024$ little boxes, iterating the mapping many
times, and counting the number of occupied boxes.  It is important
iterate the map enough times so that (a) the expected occupation number of
each box is reasonably large and (b) the trajectories have time to
wander into all the nooks and crannies.  (In practice, the second
requirement is more stringent.)  With $\xmin$, $\xmax$, and $X_r$ as
given in the caption to fig.\ \expense, the area of the stochastic
component is found to be about $b_2=0.693$.

\section \discuss.  Application of the results.

We wish now to determine the contribution of an island to the
correlation function of an orbit in the stochastic component of phase
space of a general two-dimensional area-preserving mapping $G$.  (The
analysis applies equally well to Hamiltonians with two degrees of
freedom.  The phase space is then the Poincar\'e surface of section and
the unit of time is the period of the island.)

For simplicity we begin by considering the case where there is a single
small island embedded in the connected stochastic component.  Let the
total area occupied by the stochastic component be $A_1$.  Suppose a
small island centered at $\vec x_0$ is immersed in the stochastic sea.
When iterating $\qstar$, we have been approximating $G$ in a small
region around $\vec x_0$.  The square defined by $\xmin$ and $\xmax$ in
$\qstar$ is transformed into a small box (parallelogram) of area $B_0$.
The ratio of the areas in the two spaces is $B_0/L^2=\gamma$.  Suppose
the area of the stochastic component of $G$ which lies
inside this box is $B_1$.  (In the notation of appendix \mud,
$B_1=\gamma b_1$.)  Recall that $\qstar$ also contains spurious islands
(accelerator modes) which have no counterpart in $G$.
To account for these islands we define $f_t^\ast$ as in (A1).
From this we can derive $\sum f_t^\ast =1/\alpha^\ast$
and $p_t^\ast=\alpha^\ast f_t^\ast$ (parallelling the definitions made in
section \trapping).

It is useful to begin by forming an idea of what a stochastic
trajectory will look like.  Orbits will consist of alternating trapped
and free segments.  The trapped segments are those which are restricted
to $B_1$, while the free segments are those excluded from $B_1$.  (Here
and in the following we use $B_1$, etc., to refer to a particular
subset of phase space as well as the area of this subset.)  The basic
assumption is that each visit to the island is uncorrelated with the
previous one.  So, on first entering the area $B_1$, we assume that a
segment of length $t$ will be chosen randomly with probability
$p_t^\ast$.  A simple model for the motion in the stochastic region
$A_1-B_1$, which excludes the region near the island, is as follows.  The
first point after a trapped segment is randomly (and with a uniform
distribution) situated in the $A_1-B_1$.  This is in accord with the
picture that once an orbit leaves $B_1$, it rushes away from the
vicinity of $B_1$ extremely quickly.  If this point is the pre-image
under $G$ of $B_1$, then the next point is the first point of another
trapped segment.  Otherwise another point is chosen at random in
$A_1-B_1$ and the procedure is repeated.  The mean length of
trajectories trapped in $B_1$ is $\alpha^\ast$.  The area of the points
which are initial points of trapped segments is therefore
$B_1/\alpha^\ast$.  These are the points whose pre-images lie outside
$B_1$.  Thus the probability that a point in $A_1-B_1$ is a pre-image of
$B_1$ is $\epsilon=(B_1/\alpha^\ast)/(A_1-B_1)$.  The probability that a
particular free segment has length $t$ is then $\epsilon
(1-\epsilon)^{t-1}$.  These probabilities ensure that ratio of time that
the trajectory spends in $B_1$ and in $A_1-B_1$ is in the ratio of the
area of these regions.  (The assumptions made to obtain the distribution
of lengths of the free segments is probably overly restrictive.  However
such a ``memory-less'' model is probably accurate for the long times we
are interested in.)  This model is discussed in more detail in appendix
\markov\ where it is used as a basis for constructing a Markov-chain
approximation of the motion.

Consider the correlation function
$$\Cscr(\tau)=\ave{h(\vec x(t))h(\vec x(t+\tau))}_t,\eqn(\en\ch)$$
\def\cisl{\Cscr_{\subrm{is}}(\tau)}
where $h$ is some smooth function of the position in phase space $\vec
x$ (in particular we require that it is a constant throughout $B_1$).
We shall assume that the mean value of $h(\vec x(t))$ is zero
for the stochastic orbits.  We identify those terms in (\+) for which
$\vec x(t)$ and $\vec x(t+\tau)$ belong to the same trapped segment
as $\cisl$ the contribution to $\Cscr(\tau)$ due to the island.
(Thus for such terms we have
$\vec x(t+\tau^\prime)\in B_1$ for all $\tau^\prime$ such that
$0\leq \tau^\prime\leq\tau$.)  Except for
$\tau=0$, the other terms are smaller by a factor of about $B_1/A_1$, which
is typically very small.
This is so because $h$ has a zero mean and because of the rapid mixing
of orbits in $A_1-B_1$.  This question is examined in
appendix \markov\ where it is also shown that the additional terms do not
contribute to the diffusion coefficient.  $\cisl$ may be written as
$$\cisl=h^2(\vec x_0){B_1\over A_1}
\sum_{t=\tau}^\infty (t-\tau)f_t^\ast.$$
The first factor arises because both endpoints are in $B_1$ and $h(\vec
x)\approx h(\vec x_0)$ for such points.  The second factor is the
probability that $\vec x(t)$ lies in $B_1$, and the sum
is the probability that $\vec x(t+\tau)$ belongs to the same trapped
segment as $\vec x(t)$.  This sum is just the correlation
function defined in terms of $f_t^\ast$ instead of $f_t$.
For large $\tau$, we can substitute for $f_t^\ast$ using (A2), and 
the sum
becomes $(b_2/b_1) C_\tau$.  Thus the contribution to the correlation
function due to the island is
$$\cisl=h^2(\vec x_0) {\gamma\over A_1}b_2 C_\tau.$$
Equation (A3) shows that this result is independent (for large $\tau$) of
the choice of $\xmax$ and $\xmin$ (which is as it should be).

If $h$ is the rate of change of one of the components of
$\vec x$, e.g., $h(\vec x)=dx/dt$, then
the island enhances the $x$-space
diffusion coefficient by
\def\disl{\Dscr_{\subrm{is}}}
$$\disl=h^2(\vec x_0) {\gamma\over A_1}b_2 D,$$
where $D$ (assuming that it exists) is given in (\ddef).
Whether or not $D$ exists, the mean squared $x$ position
of a group of particles
initially concentrated in a small region is enhanced by
\def\sisl{\Sscr_{\rm is}(t)}
$$\sisl=h^2(\vec x_0) {\gamma\over A_1}b_2 S_t,$$
where $S_t$ is given by (\spreaddef).

If there are more than one island, then the contributions
should be added together in $\Cscr_\tau$ and $\Dscr$.  If there
is a chain of $N$th order islands
at $\vec x_0$, $\vec x_1$, $\ldotss$, $\vec x_N=\vec x_0$, their
contribution to the correlation function $\Cscr(N\tau+j)$ is
$$\Cscr_{\subrm{is}}(N\tau+j)=
\sum_{i=0}^{N-1} h(\vec x_i)h(\vec x_{i+j}) {\gamma\over A_1}b_2 C_\tau,$$
where $\gamma=B_0/L^2$ for one of the islands and $0\le j<N$.  The
contribution to $\Dscr$ (assuming again that it exists) is
$$\disl=
\bigglp\sum_{i=0}^{N-1} h(\vec x_i)\biggrp^2 {\gamma\over A_1}b_2 D,$$
and similarly for $\sisl$

Let us apply these results to the first-order accelerator modes for the
standard mapping (\sm).  These are first-order islands which appear at
$k=2\pi n$ ($n$ an integer) \ref\chirikov.  The acceleration in such a
mode is $r_t-r_{t-1}=\pm2\pi n$.  These modes appear in pairs, one with
each sign of the acceleration.  We will take the area of the stochastic
component $A_1$ to be equal to the entire area of phase space $(2\pi)^2$.  The
relation between the parameters is found by matching the residues at
the stable fixed point.  This gives $k^2 = (2\pi n)^2+16K$.  When
transforming to the quadratic map $Q$, lengths are magnified by a
factor $\half\pi n$ \ref\krw\ and so $\gamma=(2/\pi n)^2$.  If we wish
to estimate the diffusion coefficient, we must define $h$ to be the
acceleration; thus $h(\vec x_0)=\pm 2\pi n$.  For $K=0.1$, we take
$b_2=0.693$ and the proportionality constant connecting the scripted and
unscripted quantities above is $2h^2(\vec x_0) (\gamma/ A_1)b_2 
\approx0.56$.  (The factor of 2 accounts for the presence of the
two islands.)

Using $\Dscr_{\subrm{ql}}=(\pi n)^2$ and taking $6400$ as a lower
bound for $D$,
we find that the contribution to the diffusion coefficient is
increased over its quasi-linear value by a factor of at least
$360/n^2$.
Thus for $n=1$ or $k\approx 6.41$, the islands completely dominate the
diffusion.  The first-order accelerator modes continue to have such a
large effect at least until $k\approx 100$.
If $D$ is in fact infinite, even arbitrarily small accelerator
modes will eventually dominate the motion and
fig.\ \spread\ can be used to estimate
the time at which the accelerator modes become important.

\section \conclusions.  Discussion.

We have looked at the effect of a small island on the correlation
function for the stochastic trajectories of Hamiltonians of two degrees
of freedom.  When the parameter $K$ is small, the problem may be treated
analytically and we find that the contribution to the correlation
function $C_\tau$ is zero for $\tau\grapprox 5.1454 \abs K^{-1/4}$ when
$K<0$, decays as $\tau^{-5}$ when $K=0$, and decays as
$\exp(-K^{-1/4}\tau)$ for $K>0$.

The more interesting case is when $K$ is not small and the island is
surrounded by other islands.  In the case we considered in detail
$K=0.1$, the decay of the correlation function is algebraic and very
slow (roughly as $\tau^{-1}$) out to times on the order of $\tau\sim 10^7$.
Even when the islands are small, this can still lead to an enormous
increases of quantities such as the diffusion coefficient.  Although
it has not been definitely established, there are strong indications
that the diffusion coefficient may be infinite, indicating that
the distribution of particles does not obey a diffusion equation and
that the particles spread more rapidly than diffusively.  Such
behavior is fairly typical, having been observed at several other
values of $K$.

It would be interesting to know how the distribution of particles
does evolve in time.  If we consider
a system like the standard mapping at a parameter value for which an
accelerator mode exists, then for large $t$ this distribution could, in
principle, be found from $f_t$, but its determination is beyond the
scope of this work.  For now, we observe that the distribution will be
far from Gaussian for times at least until $t=10^8$.  It will
contain a very small but extremely long tail that contributes
significantly to the variance.  This has important consequences for
numerical experiments.  Imagine trying to measure $S_t$ for
$t=10^8$ by directly measuring $r_t-r_0$ for $N$ particles.  If $N$ is
merely some ``reasonably'' large number like $1000$ then $S_t$ will
most likely be greatly underestimated.  (To counteract this there is a
tiny probability that $S_t$ will be fabulously overestimated.)  We know
that we should sample some orbits with $r_t-r_0\sim 10^7$ to be able to
estimate $S_t$ accurately.  But since $S_t\sim 10^4$, we need to sample
at least $(10^7)^2/10^4=10^{10}$
orbits before the effect of such a long segment is
correctly diluted.  Obviously such a calculation is totally out of the
question.  A much better approach is to measure the correlation
function and to derive $S_t$ using (\spreaddef).  On the
other hand, the requisite number of orbits probably are sampled
in real experiments.  For instance in plasma physics applications the
total number of orbits is typically $10^{14}$.

There are several questions still to be answered.  What is the long-time
behavior of the correlation function?  Figure \expense\ shows that it
has not attained any well-defined asymptotic limit by $\tau=10^7$.  What
determines the asymptotic behavior of the correlation function?  The
simplest picture we can form for treating this problem would go
something like this:  There is some outer KAM curve marking the boundary of
the main island centered at $(-\sqrt K,0)$.  According to
Greene\ref\greene, this curve is approximated from the outside by a
sequence of islands whose winding numbers are the rational approximants
to the irrational winding number of the KAM curve.  Thus the long time
behavior of $f_t$ may be found by considering how an orbit wanders
through these islands to approach the KAM curve.  Greene
\ref\greenepriv\ found an algebraic decay of the
correlation function based on a simple model of such a process.
Equivalently this behavior
may be obtained from a diffusion equation in which the diffusion
coefficient approaches zero sufficiently fast as the KAM curve is
approached.  This picture was the one proposed in ref.\ \chirikova.

Such pictures are
however probably incomplete.  There is no reason to suppose that the KAM
curve around the central island determines the asymptotic behavior of
$f_t$.  For $K=0.1$, it could equally well be the last KAM curve around
the sixth-order islands which surround the central island, or the KAM
curve around one of the chains of islands around the sixth-order
islands, or all of them together!  For instance,
the longest orbit segment seen for $K=0.1$, whose
length was about $6\times10^7$, spends most of its time around such a
chain of islands whose order is $6\times23=138$.  (One of the members of
this chain is visible in fig.\ \exa(b) at $x+\sqrt K\approx0.352$ and
$y\approx 0.034$.  An enlargement of this long orbit is shown in fig.\
\long.)  Is it the KAM surface around this island chain that
will determine the asymptotic behavior of $f_t$?  In this picture, the
asymptotic behavior is determined by the islands at a finite depth in
the islands-around-islands hierarchy.  This may be false.  Perhaps as $t$
is increased, we must look at deeper and deeper levels of the
hierarchy.  Such is the view taken by Chirikov in ref.\ \chirikovb.
We may also have to jump between branches of the hierarchy
as $t$ increases.  (This is supported by the observation that the three
longest orbits seen for $K=0.1$ are each stuck close to different island
chains.)

The absence of any idea as to the asymptotic behavior of $C_\tau$ makes
it virtually impossible either to obtain numerically an upper bound on the
diffusion coefficient or to establish that it is unbounded.  Such
questions can probably only be answered when this problem is better
understood analytically.  In any case, it may
be necessary to take a critical look at
extraneous effects which act as an extrinsic noise source and which
will cut off the very long correlations.

Finally, what is a measure of these islands?  This gives the overall
importance of this phenomenon.  We already have Sinai's estimate as to
the number of islands that will be created.  What is still needed is
some estimate of their size and of the interval in parameter space in
which they exist.  Numerical evidence suggests that these both decrease
rapidly with the period.  The latter quantity would enable us to say what is
the measure of parameter space on which we expect to see islands.

\acknowledgments

This work was supported by the U.S. Department of Energy under Contract
DE--AC02--76--CHO3073.  I began it while at the Institute of Plasma
Physics, Nagoya University, participating in the U.S.-Japan Fusion
Cooperation Program.  Some of the work was carried out while at the
Aspen Center for Physics.  I would like to thank J. M. Greene, R. S.
MacKay, and F. Vivaldi for stimulating discussions.  B. V. Chirikov
provided some enlightening comments on a preliminary version of this
paper.

\def\secflag{A}\appendix \mud.  Properties of the periodic mapping.

In order to obtain correct trapping statistics, we would like to
uniformly sample the trajectories the area outside the islands of $Q$
and measure how long each trajectory spends close to the islands.  We
achieve this goal by looking at long trajectories of $\qstar$.  In this
case, ``close'' is determined by $\xmin$ and $\xmax$.  However, fig.\
\exa\ shows that the area outside the islands is not sampled uniformly.
There are holes in it corresponding to the spurious islands.  Suppose
that the order of a particular set of islands is $n$.  Particles in this
set of islands will have looped around the torus at least once in the
$x$ or $y$ direction after $n$ iterations of $\qstar$.  (These islands
are accelerator modes.)  A trajectory of length $n$ within these islands
is made up of one or more segments.  It cannot be part of a longer
segment.  By omitting the trajectories within the islands we are
therefore undercounting the segments with lengths less than or equal to
$n$.  

We can express this in a more quantitative way.  The total area of phase
space in $\qstar$ is $b_0=L^2$.  We define $b_1$ to be that part of
$b_0$ which is not occupied by the islands of $Q$ and $b_2$ to be the
area of the stochastic component of $\qstar$, i.e., that part of $b_0$
which is not occupied by the islands of $\qstar$.  The difference
between these two, $b_1-b_2$ gives the area of the spurious islands.
To include the
effect of the spurious islands, we define $f_t^\ast$ by
$$f_t^\ast={b_2\over b_1}f_t+{b_1-b_2\over b_1} g_t,\eqn(\e)$$
where $g_t$ is the contribution from the regions of phase space
occupied by the spurious islands.  If $g_t$ is normalized like $f_t$ to
have a first moment of unity, then $f_t^\ast$ has the same
normalization.  The factors multiplying $f_t$ and $g_t$ are just the
relative areas of those regions of phase to which these trapping
statistics apply.  Now $g_t$ specifies the lengths of orbit segments in
the spurious islands, which must be less than or equal to $n_{\subrm{max}}$
the maximum period length for these islands.  Thus $g_t$ is zero for
$t>n_{\subrm{max}}$.  If we wish to calculate $C_\tau$, we need only
know the trapping statistics for $t>\tau$.  So as long as
$\tau>n_{\subrm{max}}$, we may ignore $g_t$ and use
$$f_t^\ast={b_2\over b_1}f_t.\eqn(\e)$$

These spurious islands have one other interesting effect.  They
introduce correlations between the lengths of successive segments.
Consider an orbit close to such an island chain of period 3.  Then, in the
simplest case, we expect to see a succession of segments of length 3.
Such correlations are obviously a peculiarity of $\qstar$.  In the
general mapping, where the particle spends some time in the stochastic
sea before returning to the neighborhood covered by $\qstar$, the
trapped segments will be sampled randomly.

We next examine the effect of the definition of ``close,'' i.e., at the
effect of changing $\xmin$ and $\xmax$.  Consider the effect of
increasing the box size by decreasing $\xmin$ and increasing $\xmax$.
This has two effects:  new segments which never entered the original box
appear; and all the original segments may be extended both forwards and
backwards.  As long as the original box was large enough, the new
segments will all be short (since they never get close to the islands).
By the same argument, the extension of the orbit segments will be
small.  Because of the extremely rapid departure of orbits away from the
islands, we can, for instance double, the edge of the box and only
increase the average length of the orbits segments by less than one.  So
the main effect of increasing the box size will be to increase the
number of short segments.  As in the case of the accelerator modes, we
can quantify this.  If we change $\xmin$ to $\xmin^\prime$ and $\xmax$
to $\xmax^\prime$, then, for large enough $t$, the new trapping
statistics $f_t^\prime$ are given by
$$b_2^\prime f_t^\prime=b_2 f_t,\eqn(\e)$$
where $b_2^\prime$ is the area of the stochastic component in the new
box.

Lastly, we address the problem of estimating the trapping statistics
from an orbit of length $T$.  Consider an infinitely long time record
consisting of segments of various lengths $t=1,2,\ldots$.  
Pick a random record of length $T$ and call $N_t$ the expected number
of segments of length $t$.  (We do not count partial segments at the
ends.)  Then
$$N_t=\case{0&\for t>T,\cr (T+1-t)f_t&\for 0<t\leq T.\cr}$$
This may be proved by induction.  When $T=0$, we have $N_t=0$ as
required.  When we extend the record from $T-1$ to $T$, the only way a
new segment of length $t$ can be included
is if the particle is at the end of such a
segment at time $T$.  Because the probability of this happening is $f_t$,
$N_t$ increases by $f_t$ for $t\leq T$.

Thus we divide actual number of orbit segments by $(T+1-t)$ to
obtain an unbiased estimate of the trapping statistics,
$f_t$.  The proof given
above does not depend on the segments appearing in a random order.  This
is important because, as we have seen, there will be correlations between
successive segments in $\qstar$.
One problem with the method we use for measuring
$N_t$ is that the beginning of the record is not arbitrary, because
initially we have $(X,Y)$ uniformly distributed on the line $X=0$.  If
$T$ is large, this is not expected to affect the results very much.

\appendix \numer.  Numerical methods used.

In order to make the reduction of $x$ and $y$ in $\qstar$ to their base
intervals as cheap as possible, the $L\times L$ square is reduced to a
unit square with a zero origin.  Thus new coordinates $X$ and $Y$ are
defined by
$$x=\xmin+LX,\qquad y=-\half L+LY.$$
In these coordinates $\qstar$ becomes
$$N:\qquad Y_t=Y_{t-1}+G(X_{t-1})\mod1,
    \qquad X_t=X_{t-1}+Y_t-\half\mod1,$$
where
$$G(X)=aX^2-bX+c,\qquad a=2L,\quad b=-4\xmin,\quad c=2(\xmin^2-K)/L.$$
We will always choose $\xmin<-\sqrt K$, so that the stable fixed point
at $(x,y)=(-\sqrt K,0)$ is included in the base intervals.  Thus the
quantities $a$, $b$, and $c$ are all positive.

$N$ is implemented using fixed-point arithmetic on the Cray--1.  This
should be compared with the use of integer mappings by
Rannou\ref\rannou.  In both cases ``exact'' (in a sense to be defined
later) numerical mappings can be defined.  The principal difference is
in the coarseness of the grid on which the mapping is defined.  The
finest grid that Rannou used was $800\times800$.  In contrast, our grid
is about $2^{48}\times2^{48}$, so that its resolution is close to that
obtained with floating point numbers.  The use of fixed-point
arithmetic also results in faster performance because the extraction of the
integer and fractional parts of a number just correspond to masking
operations.

First we define the notation for fixed-point numbers.  We view the
full 64-bit word as representing a number in twos-complement binary
notation with the ``binary'' point placed before the last 48 bits.
Addition and subtraction are performed by the integer add and subtract
instructions.  Multiplication of fixed-point numbers is accomplished by
the floating-point multiply instruction provided the two numbers lie in
the range $[0,1)$.  Multiplication of a number by $2^m$
where $m$ is a non-negative integer is accomplished by shifting the
word to the left $m$ places (with zeros introduced on the right).

The function $G$ is then computed by writing it as
$$G(X)=2^m(AX^2-BX+C),\qquad (A,B,C)=2^{-m}(a,b,c)$$
where $m$ is the smallest non-negative integer such that $A$, $B$, and
$C$ all lie in $[0,1)$.  $X$ also lies in this range so that all the
multiplications in $G$ may be carried out by computing $AX^2-BX+C$ with
the add, subtract, and multiply instructions and shifting the result $m$
places to the left.  Since the low $m$ bits of $G$ are always zero, then,
assuming that the low $m$ bits of both $X$ and $Y$ are initially zero,
they will remain zero.  Thus there will be $n=(2^{48-m})^2$ points
accessible in the unit square.  (Typically $m=2$ and $n\approx10^{27}$.)

We are now ready to give the procedure for gathering the trapping
statistics $f_t$.
\def\item{\par\noindent\hangindent 20pt }
\item (1) Read in $K$, $\xmin$, $\xmax$, $X_r$, $T$, $q$.  (The
trapping statistics will be collected for the $q$th power of the
mapping.  $T$ is the total number of iterations of $N^{-q}$.  $X_r$ defines
the size of the randomizing zone.)
\item (2) Initialize $B_i\from0$ for $0<i\leq 150$,
$t\from0$, $\tau\from T$, $(X,Y)\from(0,0)$.
Initialize the random number generator with a ``random'' seed.  ($B_i$
are the bins used for collected the trapping statistics.  The counter
$t$ gives the length of the orbit segment so far.
The number of steps still to do is given by $\tau$.)
\item (3) Set $L\from\xmax-\xmin$, $a\from2L$, $b\from-4\xmin$,
$c\from2(\xmin^2-K)/L$.  Calculate $m$, $A$, $B$, $C$, according to the
the prescription given above.
\item (4) Set $\tau\from\tau-1$.  If $\tau<0$ stop.
\item (5) Initialize $P\from\hbox{false}$.  ($P$ is set to true
when the trajectory leaves the base square.)  Set $t\from t+1$.
\item (6) Set $(X,Y)\from R(X,Y)$ where the mapping $R$ is given
below.
\item (7) Set $(X,Y)\from N^{-q}(X,Y)$ where the mapping $N^{-q}$
is given by iterating the mapping $N^{-1}$ given below $q$ times.
\item (8) If $P$ is false go to step 4.
\item (9) Record new orbit segment by setting $i\from b(t)$,
$B_i\from B_i+1$.  Re-initialize $t\from 0$.  Go to step 4.

\noindent The mapping $R$ is defined by
\item (6.1) If $X>X_r$ return.
\item (6.2) Set $P\from\hbox{true}$, $Y\from Y+U$ where $U$ is a
uniformly distributed random variable $\mod1$.  Return.

\noindent The mapping $N^{-1}$ is defined by
\item (7.1) Set $X\from X-Y+\half$.  If $\lfloor X\rfloor\neq0$, set
$P\from \hbox{true}$.  Set $X\from X-\lfloor X\rfloor$.
($\lfloor X\rfloor$ is the largest integer satisfying
$X\ge\lfloor X\rfloor$.)
\item (7.2) Set $Y\from Y-G(X)$ where $G(X)$ is calculated as outlined
above.  If $\lfloor Y\rfloor\neq0$, set $P\from \hbox{true}$.
Set $Y\from Y-\lfloor Y\rfloor$.  Return.

\noindent The bin number $b(t)$ is defined by
$$b(t)=\case{t&\for t\leq50,\cr 
\lfloor30\log_{10}t\rfloor&\for 50<t<10^5,\cr
150&\for 10^5\leq t.\cr}$$
Finally the trapping statistics $f_t$ are obtained by
$$F_i=\case{B_i/(T+1-i)&\for i\leq50,\cr 
{B_i/(T+1-10^{(i+1/2)/30})\over
\lceil 10^{(i+1)/30}\rceil - \lceil 10^{i/30}\rceil}
&\hbox{otherwise},\cr}$$
together with
$$f_t=F_i/q^2,$$
where $i=b(\lfloor t/q\rfloor)$.

The mapping $R$ randomizes the $Y$ positions of the particles in a thin
zone ($0\leq X\leq X_r$) on the left edge of the square.  ($X_r$ is
chosen to be small, typically 0.02.)  As long as the randomizing zone is
sufficiently far removed from the islands, $R$ has no effect on the
length of orbits which are trapped for moderately long times.  $R$
serves two purposes.  When it is first applied, it picks a random
initial condition on the line $X=0$.  It also stops the trajectory from
being periodic either because of the spurious accelerator modes or
because a stochastic trajectory has eventually returned to its starting
position.  This results in a more rapid sampling of phase space and so
better statistics are obtained.

Since the accumulation of the statistics $B_i$
is quite expensive compared with iterating the map, we look
instead at the $q$th power of the mapping.  We count the orbit segment
as having ended if the orbit wrapped around the torus in any of the $q$
iterations.  This leads to some uncertainty in $f_t$ for $t\sim q$.  However,
we are most interested in very long trapping times for which $t\grgr
q$.

Full advantage is taken of the vectorization capabilities of the
Cray--1 by following $M$ (a multiple of 64) orbits and averaging the
results.  Finally, the critical
parts of the algorithm were handed-coded in {\sc Cal}, the Cray
Assembly Language.  The execution time is
approximately $(360+73q)T\ \unit{ns}$ per particle.  This is about 6
times faster than a straightforward implementation in {\sc Fortran} using
floating-point arithmetic.  The most time-consuming computations were
those of fig.\ \expense\ where $M=1600$ orbits of length
$qT=2\times 10^9$ were used, a total of $3.2\times10^{12}$
iterations of $N$.  This consumed about $65$ hours of {\sc cpu} time on the
Cray--1.

Aside from the computation of $G(X)$, all the calculations are exact.
$N$ is therefore area-preserving, or, since only a discrete set of
numbers may be represented on the computer, it is better to say
that it is one-to-one.  A trajectory may be run backwards by
$$N^{-t}=J^{-1}N^tJ$$
where $J$ is the involution ($J^2$ is the identity)
$$J:\qquad X_t=X_{t-1}-Y_{t-1}+\half \mod1,\qquad
Y_t=-Y_{t-1} \mod1$$
which may be calculated exactly using fixed-point arithmetic.

Because of the discreteness of the number system on a computer, all the
trajectories of a numerical mapping are eventually periodic.  Because
$N$ is a one-to-one mapping, the whole trajectory will be periodic.
There will be no initial transient.  Presumably the typical period
length of an orbit in the stochastic part of phase space is very long.
Rannou\ref\rannou\ gives $\half (n+1)$ as the average length of a
trajectory in a random permutation of $n$ points.  (This is to be
compared with the period length for an orbit in a random {\it function}
on $n$ points which is $\sqrt{\pi n/8}+\fract1/3$
approximately\ref\knuth.  The distinction between a function and a permutation is
that a permutation is a one-to-one mapping of the $n$ points onto
themselves, while a function is a many-to-one mapping of the points
into themselves.  It is not clear whether the stochastic
orbits of floating-point mappings are closer to those of random
permutations or random functions.)  However, $N$ is not a random mapping
because it possesses symmetries.  In the case of the standard mapping
this reduces the expected length of stochastic orbits to $O(\sqrt
n)$\ref\rannou.  Although this number is still large (about $10^{13}$),
we expect there to be a large number of trajectories with shorter
periods.  Examples of such trajectories are those in the accelerator
modes which are restricted to a small fraction of phase space.  By
composing the mapping with $R$, we give the trajectory a random step
whenever it returns to its starting point (and usually before that).  We
therefore ensure that a particle stays in these short periodic orbits
for at most $q$ periods.  (Of course the random number generator is
an example of a mapping.  But it is specially chosen to have a very long
period of $2^{40}$, so the random number generator repeats itself
after about $10^{12}q$ iterations of $N$.)

\appendix \markov.  Correlation function for \capital Markov model.

In section \discuss, we presented a model for the motion of a
stochastic trajectory when an island is present.  We wish to make that
model more precise so that we can accurately assess the influence of
the island on the correlation function.  We do this by modeling phase
space with a finite number of points and writing down a Markov
transition matrix $\mat P$ for evolution of the distribution function.

Suppose there are a total of $N$ points.  In the simplest case, there
is just one trapped segment of length $m$.  The stochastic sea consists
of the other $n=N-m$ points, one of which is special in that it is the
pre-image of one of the trapped points.  For example if $N=7$, $m=3$,
$n=4$, then $\mat P$ is given by
$$\mat P={\def\c{\fract1/4}\matrix7{
0 &0 &0 &0 &0 &0 &1 \cr
1 &0 &0 &0 &0 &0 &0 \cr
0 &1 &0 &0 &0 &0 &0 \cr
0 &0 &\c&\c&\c&\c&0 \cr
0 &0 &\c&\c&\c&\c&0 \cr
0 &0 &\c&\c&\c&\c&0 \cr
0 &0 &\c&\c&\c&\c&0 \cr}}.\eqn(\en\pex)$$
The entry in the $j$th row and $i$th column is the probability that a
particle at point $i$ at time $t$ is at point $j$ at time $t+1$.  Thus
we see that points 1--3 constitute the trapped segment, while points
4--8 are the stochastic points with point 8 being the pre-image of one
of the trapped points.

\mathchardef\bunion="225B\mathchardef\binter="225C
In the more general case, there may be $M$ trapped segments, whose
lengths are $m_k$ for $1\leq k\leq M$.  Suppose the members of the $k$th
trapped segment are $T_k=\{p_k^{(1)},p_k^{(2)},\ldotss,\penalty10p_k^{(m_k)}\}$,
and that the pre-image of $p_k^{(1)}$ is $q_k$.  We will also
define $T_k^\prime= \{q_k\}\bunion T_k$, $T=\union_k T_k$,
$T^\prime=\union_k T_k^\prime$, $S=T^{\subrm c}$,
$S^\prime=T^{\prime\subrm c}$, where the superscript c denotes the
complement, and $m=\sum_k m_k$.  Thus $T$ is the set of trapped points,
$T^\prime$ is $T$ with the pre-image points added, $S$ is the set of
stochastic points, $S^\prime$ is the $S$ with the pre-image points
removed, and $m$ is the total number of trapped points.  It is easy
to generalize (\pex) to incorporate the additional trapped
segments.

If $\vec F(t)$ is a column vector giving the distribution function at
time $t$, then
$$\vec F(t)=\mat P^t\vec F(0).$$
As $t\to\infty$, $\mat P^t$ approaches a matrix $\mat C$ each of whose entries is
the constant $1/N$.  It is convenient to define $\mat Q$ such
that $\mat Q=\mat P-\mat C$.  It is easy to show that, for $t>0$,
$\mat Q^t=\mat P^t-\mat C$ so that $\lim_{t\to\infty}\mat Q^t=\mat 0$.
Since so many of the entries in $\mat Q^t$ are the same,
the computational effort involved in doing the matrix multiplication is
the same as for multiplying $(m+1)\times (m+1)$ matrices.

The correlation
function for a function $h$ on phase space is given by
$$\Cscr(\tau)=\vec h\cdot\mat P^\tau\cdot\vec h/N,\eqn(\en\cmark)$$
where $\vec h$ is a vector giving the value of $h$ for each point.  This
parallels the definition (\ch) given in section \discuss.  As in that
section, we will assume that $\sum h_j=0$.  In that case, we have
$\vec h\cdot\mat C\cdot\vec h=0$, so that we can
replace $\mat P$ by $\mat Q$ in this definition.

Let us consider first the case where there is a single trapped segment,
$M=1$.  Suppose that $h_j$ is 1 for $j\in T$, 0 for $j=q_1$, and
takes on arbitrary values (consistent with the requirement that $\sum
h_j=0$) for $j\in S^\prime$.  In the example given in (\pex), we might
choose $\vec h=[1,1,1,-1,1,-3,0]$.  (We pick $h$ to be zero for the
pre-image point, so that there is no bias for this point.)  In the
notation of section \discuss, we have $A_1=N$, $B_1=m$,
$f_t^\ast=\delta_{tm}/m$, $\cisl=(m-\tau)/N$ for $\tau\leq m$,
$\disl=\half m^2/N$.

We have numerically determined $\Cscr(\tau)$ for this case using
(\cmark) with $m=10$, $20$, and $50$ and $N$ varying between $10m$ and
$10^6m$.  In particular, we considered $\epsilon(\tau)\eqv
(\Cscr(\tau)-\cisl)/\Cscr_{\subrm{is}}(0)$.  We found that for $\tau>0$,
$\abs{\epsilon(\tau)}\leq m/N =B_1/A_1$.  After an initial transient,
$\epsilon(\tau)$ settles down to a decaying periodic function, which
oscillates about zero, whose half-period is $m$, and which decays as
$n^{-\tau/m}$ approximately.  Thus for $N\grgr m$, the corrections to
$\cisl$ are very small.

It is, perhaps, a little surprising that this model predicts
anti-correlations where $\Cscr(\tau)<0$.  This is most pronounced for
$m\lsapprox \tau\lsapprox 2m$.  It is, however, a reflection of a real
effect in mappings.  Consider an orbit in the stochastic sea of a
general mapping.  If there is an island present, the trajectory can be
stuck close to the island for very long times.  However, because of
ergodicity, the average time the orbit spends close to the island is
equal to the relative areas of the stochastic components $B_1/A_1$.
Thus, to compensate for the stickiness, an orbit far from the island
must have some difficulty coming close to the island.  This leads to the
anti-correlations observed.  (These anti-correlations may be rather
difficult to measure directly because the presence of trapped segments
of so many different lengths will average out the oscillations in
$\Cscr$.)  A fairly accurate picture of the situation is that the
stochastic region close to the island is surrounded by a box with a
very small opening in it.  A particle inside the box has to bounce
around a lot before finding the opening and escaping.  Similarly, once
the particle is outside the box (in the main part of the stochastic
sea), it cannot easily find the opening in order to come close to the
island once again.

Because $\cisl=0$ for $\tau\geq m$, whereas $\epsilon(\tau)$ is
finite for all times, we still need to verify that the integrated
effect of $\epsilon(\tau)$ (which gives the corrections to
$\disl$) is small.  We define the diffusion coefficient in
the same way as (\ddef).  This gives
$$\Dscr=\vec h\cdot(\half\mat I+\sum_{\tau=1}^\infty
\mat Q^\tau)\cdot\vec h/N
=\vec h\cdot(\half\mat I+\mat Q\cdot(\mat I-\mat Q)^{-1})\cdot\vec h/N.
$$
We are able to write the sum in this way because $\mat Q^\tau$
converges to zero.  The
the matrix operations were performed for several examples with 1 or 2
trapped segments using {\sc Macsyma}\ref\macsyma.  The results show
that in general $\Dscr$ may be written as
$$\Dscr={1\over 2N} \bigglp\sum_{j\in S^\prime}h_j^2
+\sum_k\sum_{i\in T_k^\prime}\sum_{j\in T_k^\prime}h_i h_j\biggrp.
\eqn(\en\dmark)$$
This result is exact and only requires
that $\sum h_j=0$.  The first term in parentheses is just
the contribution from $\Cscr(0)$ for the stochastic points,
while the second term gives island contribution.
In the one-segment case discussed above, the island contribution
clearly reduces to $\half m^2/N$ which exactly
equals $\disl$.  Thus the oscillations in $\epsilon(\tau)$ are such that
its sum is precisely zero.

For the more general case, where there is an arbitrary number of
trapped segments of various lengths, there are two requirements: $h_j$
should be a constant $h_0$ for $j\in T$ and $\sum_k m_k h_{q_k}$
should be zero.  (This last condition says that the pre-image points
should be sufficiently evenly spread over phase space.)  From section
\discuss, we have $f_t^\ast=\sum_k \delta_{tm_k}/m$, $\disl=\half
h_0^2/N\sum_k m_k^2$, while from (\dmark) we obtain
$$\Dscr={1\over2N}\bigglp\sum_{j\in S}h_j^2 +
h_0^2\sum_k m_k^2\biggrp.\eqn(\e)$$
Again, the second term exactly gives $\disl$.
Alternatively, we could declare that the pre-image points are
part of the trapped segments.  (This conflicts with the
picture given in section \discuss\ because it would allow one trapped
segment to follow immediately after another with no
intervening free segment.)  Letting $h_j=h_0$ for $j\in T^\prime$
and defining $m_k^\prime=m_k+1$, (\+) would be modified by replacing
$S$ by $S^\prime$ and $m_k$ by $m_k^\prime$.

\endpaper